\documentstyle[12pt]{article}
\begin{document}
\newcommand{\bea}{\begin{eqnarray}}
\newcommand{\eea}{\end{eqnarray}}
\renewcommand{\thefootnote}{\fnsymbol{footnote}}
\newcommand{\be}{\begin{equation}}      \newcommand{\ee}{\end{equation}}
\newcommand{\st}{\scriptsize}
\newcommand{\fs}{\footnotesize}
\vspace{1.3in}
\begin{center}
\baselineskip 0.3in

{\large \bf On the genuine bound states of a non-relativistic particle in a linear finite range
potential}

\vspace {0.3in}
{ Nagalakshmi A. Rao}

\vspace{-0.2in}
{\it Department of Physics, Government Science College,}

\vspace{-0.2in}
{\it Bangalore-560001,Karnataka, India.}

\vspace{-0.2in}
{drnarao@gmail.com}
 
\vspace{0.2in}
{ B. A. Kagali }

\vspace{-0.2in}
{\it Department of Physics, Bangalore University,}

\vspace{-0.2in}
{\it Bangalore-560056, India.}

\vspace{-0.2in}
{bakagali@hotmail.com}
\end{center}

\vspace{0.3in}
\hspace{2.3in}

\begin{center}
{\bf Abstract}
\end{center}
We explore the energy spectrum of a non-relativistic particle bound in a linear finite range, attractive
potential, envisaged as a quark-confining potential. The intricate transcendental eigenvalue
equation is solved numerically to obtain the explicit eigen-energies. The linear potential, which 
resembles the triangular well, has potential significance in particle physics and exciting
applications in electronics.

\vspace{0.6in}

\vspace{0.1in}

\vspace{0.1in}
\noindent
{\bf Comments} Latex, 8 Pages, 1 Table, 1 figure.

\vspace{0.3in}
\noindent

\setcounter{section}{0}
\newpage
\indent
{\section{\large \bf Introduction}}

A challenging problem in particle physics in recent years is that of quark confinement. It is 
presently known that mesons are not elementary particles, but are composed of quarks, 
as are the nucleons. 

In literature, several approximation methods are available relating to quark confinement.
 The {\it lattice model} [1] suggests that at large distance between quarks, 
the interaction increases approximately linearly with separation. The {\it bag model}, where
 quarks and gluons are confined in a bag, is not suitable for calculating the hadronic 
properties of heavy quarks or in computing the energy levels of excited states. 
{\it String model}, on the other hand, proposes  quark-antiquark pair at the ends of an open 
string and creation of quark-antiquark pair when the string breaks. In recent years, 
{\it potential models} [2] are best justified theoretically to describe heavy quarkonia 
and seem to be most powerful in calculating the static properties.

Several authors [3 - 6] have addressed the bound states of  various kinds of 
linear potential. Chiu [7] has examined the quarkonium systems with the regulated linear plus 
Coulomb potential in momentum space. Deloff [8] has used a semi-spectral Chebyshev method for 
numerically solving integral equations and has applied the same to the quarkonium bound state
problem in momentum space. 

Rao and Kagali [9 - 11] have investigated extensively the bound states of both spin and 
spinless particles in a screened Coulomb potential, having linear behaviour near the origin.
In the present work, we propose a finite, short-ranged linearly rising potential, envisaged
 as a quark-confining potential and explore the non-relativistic bound states.

\vspace{-0.1in} 
\indent
{\section{\large \bf The Schrodinger equation with the linear potential}}
Several attempts have been made to study the meson spectra using the non-relativistic 
Schrodinger equation with a linear potential. Intuitively, we construct a simple linear rising,
 finite range potential of the form [12] 
\bea
V\left(x\right)=-{V_{0}\over a}\left(a-\left|x\right|\right),
\eea
 in which the well depth $V_{0}$ and range $2a$ are positive and adjustable parameters.
 The linear potential with its boundary regions is illustrated in Fig.1 and  
owing to its shape, this potential could also be called the trianglular potential well.

Obviously in regions I and IV, the particle is free and the allowed solutions of the free 
particle Schrodinger equation are 
\bea
\psi _{1}(x)={C}_{1}e^{\alpha x}\ \ \ -\infty <x\leq -a  
\eea 
\bea
\psi _{4}(x)={C}_{6}e^{-\alpha x}\ \ \ \ a\leq x<\infty ,
\eea
consistent with the requirement $\psi(x)$ vanishes as $\left|x\right|\rightarrow \infty $.

Here $\alpha ^{2}=-${\Large ${2mE\over \hbar {2\atop }}$} is implied. Since $E<0$ for bound states, $\alpha$ is positive. To discuss 
the nature of the soution within the potential region, $-a<x<a,$ we insert the potential 
described in Eqn.(1) in the celebrated Schrodinger equation and obtain 
\bea
{d^{2}\psi \over dx^{2}}+{2m\over {\hbar }^{2}}\left[E+V_{0}\left(1-{\left|x\right|\over a}\right)\right]\psi (x)=0 .
\eea
Setting {\Large ${x\over a}$}$=y$ and defining $\bar E=${\Large ${E\over \hbar ^{2}/2ma^{2}}$} \ and \ $\bar V_{\rm o}=${\Large 
${V_{\rm o}\over \hbar ^{2}/2ma^{2}}$}  we obtain the dimensionless form of the Schrodinger 
equation 
\bea
{d^{2}\psi (y)\over d{y}^{2}}+\left[\bar E+\bar V_{0}(1-y)\right]\psi (y)=0 ,
\eea
which may further be written as
\bea
{d^{2}\psi \over d{y}^{2}}-Ay\psi +B\psi =0.  
\eea
The constants $A=\bar V_{0 }$ and $B=\bar E+\bar {V_{0 }}$ are  also dimensionless. Introducing
 an auxiliary function 
\vspace{-0.1in}
\bea
w=A^{1\over 3}\left(y-{B\over A}\right)  \eea
yeilds
\bea
{d^{2}\psi \over dw^{2}}-w\psi =0.  
\eea
\indent
The solutions of this differential equation are the well-known Airy functions $Ai(w)$ and 
$Bi(w)$ [13], having oscillatory 
and damping nature.

\noindent
{\large \bf  The Eigenvalue Equation}

\indent
The admissible solutions in the four regions, consistent with physical reality, are 
\bea
\psi _{1}(x)=C_{1}e^{\alpha x}\ \ \ \ \ \ \ \ \ \ \ \ \ \ \ \ \ \ \ \ \ \ \ \ \ \ -\infty <x\leq -a  
\eea
\bea
\psi _{2}(x)=C_{2}Ai(w)+C_{3}Bi(w)\ \ \ \ \ \ \ \ \ -a\leq x\leq 0
\eea
\bea
\psi _{3}(x)=C_{4}Ai(-w)+C_{5}Bi(-w)\ \ \ \ \ \ \ \ 0\leq x\leq a
\eea
\bea
\ \ \psi _{4}(x)=C_{6}e^{-\alpha x}\ \ \ \ \ \ \ \ \ \ \ \ \ \ \ \ \ \ \ \ \ \ \ \ \ \ \ \ \ \ a\leq x<\infty 
\eea
where $C_{1}$ to $C_{6}$ are the normalisation constants. Imposing on the solutions in equations (9) to (12) the 
requirements that 
$\psi$ and {\Large $d\psi \over dx$} be continuous at the origin and also at the potential boundaries 
$(x=\pm a)$ leads to the eigenvalue 
eqnation.

At $x=-a,$ $\ \psi _{1}(x)=\psi _{2}(x)$ and {\Large ${d\psi _{1}\over dx}$}$=${\Large ${d\psi _{2}\over dx}$} 

This leads to 
\vspace{-0.1in}
\bea
C_{1}e^{-\alpha a}=C_{2\ }Ai(w_{1})+C_{3}\ Bi(w)  
\eea    
\bea
\alpha C_{1} e^{-\alpha a}={A^{1 \over 3}\over
a}\left[C_{2} \ Ai'(w_{1})+C_{3} \ Bi'(w_{1})\right]  
\eea
where 
\vspace{-0.1in}
\bea
{w}_{1}=A^{1\over 3}\left(-1-{B\over A}\right)
\eea
On simplification, we obtain 
\bea
\alpha ={A^{1\over 3}\over a}\left[P\ Ai'(w_{1})+Bi'(w_{1})\over P\ Ai(w_{1})+Bi(w_{1})\right]
\eea
where $P=${\Large ${C_{2}\over C_{3}}\ $}. Similarly, the continuity condition at $x=0$ demands 

$\psi _{2}(x)=\psi _{3}(x)$ \ and\ {\Large ${d\psi _{2}\over dx}$}$=${\Large ${d\psi _{3}\over dx}$}, 

from which we obtain
\bea
C_{2}Ai(w_{0 })+C_{3}Bi(w_{0 })=C_{4}Ai(-w_{0
})+C_{5}Bi(-w_{0 })
\eea   
\bea
C_{2}Ai'(w_{0 })+C_{3}Bi'(w_{0 })=C_{4}Ai'(-w_{0
})+C_{5}Bi'(-w_{0 })
\eea
with 
\vspace{-0.1in}
\bea
w_{0 }=A^{1\over 3}\left({-B\over A}\right).  
\eea
As before, 
\bea
{P\ Ai'(w_{0 })+Bi'(w_{0 })\over P\ Ai(w_{0
})+Bi(w_{0 })\ }={Q\ Ai'(-w_{0 })+Bi'(-w_{0 })\over
Q\ Ai(-w_{0 })+Bi(-w_{0 })\ }
\eea
where $Q=${\Large ${C_{4}\over C_{5}}$} is another constant. Adapting similar procedure at the boundary $x=+a,$ demanding 
$\psi_{3}(x)=\psi_{4}(x)$ and {\Large ${d\psi _{3}\over dx}$}$=${\Large ${d\psi _{4}\over dx}$} one would on similar grounds obtain 
\bea
-\alpha ={A^{1\over 3}\over a}\left[Q\ Ai'(-w_{2})+Bi'(-w_{2})\over Q\ Ai(-w_{2})+Bi(-w_{2})\right]  
\eea  
with
\bea
w_{2}=A^{1\over 3}\left(1-{B\over A}\right).
\eea
It is worthwhile mentioning that the arguments of the Airy function $w_{0},\ w_{1},\ {\rm and}\ w_{2}$ 
 are dependent 
both on the energy as well as the potential and are related by the simple equation 
\bea
w_{0}={w_{1}+w_{2}\over 2}.
\eea
It is straightforward to check that 
\bea
P={\beta \ Bi(w_{1})-A^{1\over 3}\ Bi'(w_{1})\over A^{1\over 3}\ Ai'(w_{1})-\beta \ Ai(w_{1})},
\eea
and
\bea
Q=-\left[{\beta \ Bi(-w_{2})+A^{1\over 3}\ Bi'(-w_{2})\over \beta
\ Ai(-w_{2})\ +{\ A}^{1\over 3}\ Ai'({-w}_{2})}\right],
\eea
where $\beta = \alpha a$ is implied.

Formally on eliminating $P$ and $Q$ in Eqn.(20) we obtain the eigenvalue equation as
$$\left[\left\{ \beta Bi\left(\omega _{1}\right)-A^{1\over 3}Bi'\left(\omega
_{1}\right)\right\} Ai'\left(\omega _{0}\right)+\left\{ A^{1\over
3}Ai'\left(\omega _{1}\right)-\beta Ai\left(\omega _{1}\right)\right\}
Bi'\left(\omega _{0}\right)\over \left\{ \beta Bi\left(\omega
_{1}\right)-A^{1\over 3}Bi'\left(\omega _{1}\right)\right\} Ai\left(\omega
_{0}\right)+\left\{ A^{1\over 3}Ai'\left(\omega _{1}\right)-\beta Ai\left(\omega
_{1}\right)\right\} Bi\left(\omega _{0}\right)\right]=$$ 
\bea
\left[\!\left\{ \beta Bi\left(-\omega _{2}\right)\!+\!A^{1\over
3}Bi'\left({-\omega }_{2}\right)\right\} \!Ai'\left({-\omega
}_{0}\right)\!-\!\left\{ A^{1\over 3}Ai'\left({-\omega }_{2}\right)\!+\!\beta
Ai\left({-\omega }_{2}\right)\right\} \!Bi'\left({-\omega }_{0}\right)\over
\!\left\{ \beta Bi\left({-\omega }_{2}\right)\!+\!A^{1\over 3}Bi'\left({-\omega
}_{2}\right)\right\} \!Ai\left({-\omega }_{0}\right)-\!\left\{ A^{1\over
3}Ai'\left({-\omega }_{2}\right)\!+\!\beta Ai\left({-\omega }_{2}\right)\right\}
\!Bi\left({-\omega }_{0}\right)\right]
\eea

This intricate and fairly complicated transcendental eigenvalue equation involving the Airy 
function and its derivatives is solved both graphically and numerically using
 Mathematica[14]. The real roots, which correspond to the eigenenergies, are listed in Table 1 for
a typical value of the range parameter(a). Energy $(E)$ and well-depth $(V_{0})$ are
 both expresed in units of {\Large ${\hbar ^{2}\over 2ma^{2}}$}.

\vspace{-0.1in} 
\indent
{\section{\large \bf Results and Discussion}}
One of the distinctive characteristics of quantum mechanics, in contrast to classical mechanics,
is the existence of bound states corresponding to discrete energy levels. It is 
well-known in quantum mechanics that
bound states exist for all atractive potentials, the exact number depending on the 
specific form of the potential and the dimensionality of the space.

More specifically, as is seen from the spectrum of energies listed in Table 1, for a 
finite range of the potential, deeper wells admit
excited state energies, consistent with the wisdom of quantum mechanics. 
Such studies, apart from being pedagogical in nature, are potentially exciting and significant as
it is concerned with quark confinement.

Quantum chromodynamics, which governs the quark-antiquark interaction is widely accepted 
as a good theory of strongly interacting particles. One can explore the hadronic properties 
by investigating the bound states of quarks. Our investigation concerning the linear 
potential is seeming interesting and can be regarded as a model to describe the quarkonia.

Further, the linear potential well or in other words, the triangular well has potential
 applications in electronics. Interestingly, in many semiconductor devices, it is believed 
that electrons are confined in almost triangular quantum wells [15]. 
Examples of such devices are Si MOSFETs (Metal Oxide Semiconductor Field Effect
 Transistors) widely used in digital applications and GaAs/AlGaAs MODFETs
(Modulation Doped Field Effect Transistors) used for high speed applications. 
The bound states of the linear potential is a subject of renewed interest and intensive research
 and we have extended the study of this naive potential to the relativistic domain, 
which will be reported shortly.

\indent
{\large\bf Acknowledgements}

    This work was carried out under a grant and fellowship by the University Grants Commission.

\newpage
\begin{table}[hbt]
\vspace{0.3in}
\begin {center}

{\Large \bf Table 1}

\vspace{0.3in}
{\Large \bf  Eigenenergies of a  non-relativistic particle in a linear potential}

\vspace{0.2in}
{\Large \bf $(a=1{\_ \atop }\kern-9pt\lambda )$}

\vspace{0.2in}
\begin{tabular}{|l|l|l|l|} \hline
{}&{}&{}&{}\\
{\ \ $\bar V_{0}$}&{\ \ \ \ \ \ \ $\bar E_{0}$}&{\ \ \ \ \ \ \  $\bar E_{1}$}&{\ \ \ \ \ $\bar E_{2}$} \\ 
{}&{}&{}&{}\\ \hline
\,\,\,0.01&-\,\,\,0.0000976&{}&{} \\ \hline
\,\,\,0.05&-\,\,\,0.0022242&{}&{} \\ \hline
\,\,\,0.10&-\,\,\,0.0080220&{}&{} \\ \hline
\,\,\,0.20&-\,\,\,0.0268461&{}&{} \\ \hline
\,\,\,0.30&-\,\,\,0.0519188&{}&{} \\ \hline
\,\,\,0.40&-\,\,\,0.0808526&{}&{} \\ \hline
\,\,\,0.50&-\,\,\,0.1122377&{}&{} \\ \hline
\,\,\,0.60&-\,\,\,0.1451770&{}&{} \\ \hline
\,\,\,0.70&-\,\,\,0.1790694&{}&{} \\ \hline
\,\,\,0.80&-\,\,\,0.2134967&{}&{} \\ \hline
\,\,\,0.90&-\,\,\,0.2480162&{}&{} \\ \hline
\,\,\,1.0&-\,\,\,0.2828516&{}&{} \\ \hline
\,\,\,2.0&-\,\,\,0.6121732&{}&{} \\ \hline
\,\,\,3.0&-\,\,\,0.9072505&{}&{} \\ \hline
\,\,\,4.0&-\,\,\,1.2056918&{}&{} \\ \hline
\,\,\,4.28&-\,\,\,1.2984370&{-\,\,\,0.0000991}&{} \\ \hline
\,\,\,5.0&-\,\,\,1.5770587&{-\,\,\,0.1653382}&{} \\ \hline
10.0&-\,\,\,5.8335771&{-\,\,\,1.3641773}&{} \\ \hline
20.0&-16.4420518&-\,\,\,2.3126501&{} \\ \hline
20.62&-17.0607820&-\,\,\,2.3662130&{-0.0000343} \\ \hline
25.0&-21.2732190&-\,\,\,3.6112921&{-2.4865189} \\ \hline
30.0&-25.7497380&-\,\,\,9.0650370&-2.9501500 \\ \hline
35.0&-29.8836000&-15.0604510&-3.2815680 \\ \hline
40.0&-33.7181130&-21.1744464&-3.5904940 \\ \hline
\end{tabular}
\end {center}
\end{table}

\newpage
\baselineskip 0.2in
{\large\bf References}
\newline
\newline $[1]$ Kogut J and Susskind L 1975 { Hamiltonian formulation of Wilson's lattice guage
theories} {\it Phys.Rev D} {\bf 11} 395
\newline
\newline $[2]$ Lichtenberg D B 1987 { Energy levels of quarkonia in potential models} 
{\it Int.J.Mod.Phys. A} {\bf 2} 1669
\newline
\newline $[3]$ Antippa A F and Phares A J 1978 {The linear potential: A solution in terms of 
combinatorics functions} {\it J.Math.Phys} {\bf 19} 308
\newline
\newline$[4]$ Antippa A F and Toan N K 1979 {The linear potential eigen energy equation I}
 {\it Can.J.Phys.} {\bf 57} 417
\newline
\newline$[5]$ Plante G and Antippa A F 2005 {Analytic solution of the Schrodinger equation for
 the Coulomb plus linear potential - The wave functions} {\it J.Math.Phys.} {\bf 46} 062108
\newline
\newline$[6]$ Antonio de Castro 2003 Bound states by a pseudoscalar Coulomb potential in one plus one
dimension arXiv:hepth/0303 175v2
\newline
\newline$[7]$ Chiu T W 1986 {Non-relativistic bound state problems in momentum space} 
{\it J.Phys.A Math.Gen.}{\bf 19} 2537
\newline
\newline$[8]$ Deloff A 2007 {Quarkonium bound state problem in momentum space revisted}
{\it Ann.Phys.} {\bf 322} 2315
\newline
\newline$[9]$ Nagalakshmi A Rao and Kagali B A  2002 {Spinless particles in a screened Coulomb 
potential} {\it Phys.Lett. A} {\bf 296} 192
\newline
\newline$[10]$ Nagalakshmi A Rao and Kagali B A  2002 {Dirac bound states in a one-dimensional 
scalar screened Coulomb potential} {\it Mod.Phys.Lett. A} {\bf 17} 2049
\newline
\newline$[11]$ Nagalakshmi A Rao and Kagali B A  2002 {Bound states of Klein-Gordon particles
in scalar screened Coulomb potential} {\it Int.J.Mod.Phys.A} {\bf 17} 4793
\newline
\newline$[12]$ Nagalakshmi A Rao 1996 {\it A study of bound states in relativistic quantum
mechanics} M. Phil Dissertation (Bangalore University)
\newline
\newline$[13]$ Abramowitz M and stegun I A 1965 {\it Handbook of Mathematical Functions and
Formulas, Graphs and Mathematical Tables} (New York: Dover)
\newline
\newline$[14]$Wolfram S 1996 {\it The Mathematica Book} (Cambridge: Cambridge University Press)
\newline
\newline$[15]$ Jasprit Singh 1997 {\it Quantum Mechanics - Fundamentals and Applications to
 Technology} (New York: Wiley Interscience)
\newline

\end{document}